\documentstyle[sprocl,epsfig]{article}

\def\nostrocostrutto#1\over#2{\mathrel{\mathop{\kern 0pt \rlap 
  {\raise.2ex\hbox{$#1$}}}
  \lower.9ex\hbox{\kern-.190em $#2$}}}

\newcommand{\be}{\begin{equation}}
\newcommand{\ee}{\end{equation}}
\newcommand{\ba}{\begin{eqnarray}}
\newcommand{\ea}{\end{eqnarray}}

\begin{document}
\pagestyle{empty}

\null

{\large 

\rightline{MPI-PhT/96-91} 
\rightline{October 1996} 
\vspace{2cm}

 \centerline{\LARGE\bf The soft limit of the energy 
 spectra in QCD jets\footnote{\normalsize 
 to be published in the Proceedings of the 
 XXVI International Symposium on Multiparticle Dynamics, 
 Faro, Portugal, September 1$^{st}$-5$^{th}$, 1996} }

\vspace{0.6cm} 

\centerline{SERGIO LUPIA}

\vspace{0.6cm}

\centerline{\it Max-Planck-Institut f\"ur Physik} 
\centerline{\it (Werner-Heisenberg-Institut)} 
\centerline{\it F\"ohringer Ring 6, D-80805 M\"unchen, Germany}
\centerline{E-mail: lupia@mppmu.mpg.de}

\vspace{1.5cm} 
\centerline{\bf Abstract} 
\bigskip 

\noindent 
A scaling law for the one-particle invariant density $E 
\frac{dn}{d^3p}$ at small momenta is observed in experimental data. 
We show that these results are consistent with the predictions of 
the analytical QCD approach, based on 
Modified Leading Log Approximation  
plus Local Parton Hadron Duality, which includes 
colour coherence in  soft gluon bremsstrahlung.

\newpage 

 \centerline{\Large\bf The soft limit of the energy  spectra in QCD jets} 
 \vspace{0.5cm} 

\centerline{\large Sergio Lupia}

\vspace{0.5cm}

\centerline{\it Max-Planck-Institut f\"ur Physik} 
\centerline{\it (Werner-Heisenberg-Institut)} 
\centerline{\it F\"ohringer Ring 6, D-80805 M\"unchen, Germany}
\centerline{E-mail: lupia@mppmu.mpg.de}

\bigskip


\begin{abstract}
A scaling law for the one-particle invariant density $E 
\frac{dn}{d^3p}$ at small momenta is observed in experimental data. 
We show that these results are consistent with the predictions of 
the analytical QCD approach, based on 
Modified Leading Log Approximation  
plus Local Parton Hadron Duality, which includes 
colour coherence in  soft gluon bremsstrahlung.  
\end{abstract}


\section{Introduction}

The aim of our research project is to test the validity of the 
perturbative QCD approach to multiparticle production in the 
semihard and soft region; we also wish to study the
sensitivity of the physical observables to different aspects included in the
theory, like the running of the coupling $\alpha_s$ (see in this respect 
\cite{lo}); in this paper we mainly concentrate on the sensitivity 
of soft particle production to QCD coherence. 

The theoretical framework for the description 
of inclusive observables in jet physics at parton level is provided by 
the Modified Leading Log Approximation (MLLA) 
of QCD (for a review see~\cite{DKMTbook}), 
where coherence, running of $\alpha_s$  and 
 energy-momentum conservation are taken into account. 
Predictions depend on two free parameters only, 
 i.e., the infrared cut-off at which the parton evolution is stopped, $Q_0$, 
 and the effective QCD scale $\Lambda$ which appears in the 
 one-loop expression for the running coupling. An interesting limiting case is
 obtained by pushing the parton cascade down to the very end, i.e., by 
 choosing $Q_0 = \Lambda$; in this case, one obtains a simple closed expression for the
 spectrum, called Limiting Spectrum. 
To connect predictions at parton level with experimental data, 
Local Parton Hadron Duality (LPHD)\cite{LPHD} is taken 
as hadronization prescription, i.e., the 
inclusive hadron spectra  are required to be 
proportional to the corresponding inclusive parton spectra. 
The whole hadronization is then parametrized in terms of only one parameter,
 which gives the overall normalization of the distribution, but does not
 affect its moments of order greater or equal than one.

\section{Phenomenology of inclusive energy spectra: an update} 

The description of the inclusive energy spectrum  for charged particles 
is one of the main successes of the analytical QCD approach\cite{DKMTbook}; 
this approach turned out to be valid at LEP-1.5 $cms$ energy
too. Let us just list a few points of interest (see \cite{klo} for details). 

\noindent{\it Analysis of the shape.} 
The inclusive energy distribution is 
well described by the Limiting Spectrum 
with $Q_0$ = $\Lambda$ = 270 MeV and 3 active flavours. 
After a rescaling, which allows to have the same kinematics both at parton and
hadron level, the very soft tail of the distribution is well described as well.
The overall normalization factor at LEP-1.5 
turns out to be consistent with the LEP-1 value, in agreement with the
predictions of  the perturbative approach. 

\noindent {\it The position of the maximum.} 
The energy dependence of the position of the maximum of the spectrum is 
well described by the Limiting Spectrum predictions
with $Q_0$ = $\Lambda$ = 270 MeV and 3 active flavours. 
Notice that the popular three terms formula~\cite{DKMTbook} for the energy dependence
of the maximum underestimates by an almost constant value 0.1 the true
position of the maximum of the Limiting Spectrum; this gives rise to a 
difference of the  order of 20-30 MeV 
in the determination of  the best value of the cut-off $Q_0$. 
Let us also point out  
that the number of flavours enters in the expressions for the cumulant moments 
through the running coupling $\alpha_s(y,n_f)$; since the MLLA is defined at
one-loop level, a scale ambiguity in the  expression of $\alpha_s$ is present; 
 kinematical reasons\cite{klo} suggest to push the heavy quark 
thresholds to larger scales and keep 3 active flavours only.

\noindent {\it Moment analysis.} 
This study  is particularly
interesting, because  it does not depend on the overall normalization 
parameter and, since theoretical
 predictions  are in this case absolutely normalized at threshold, it 
 allows  to test the perturbative picture down to low $cms$ energies. 
Theoretical predictions  of the Limiting Spectrum with 
$Q_0$ = 270 MeV and 3 active flavours  are in good  agreement 
with experimental results\cite{lo}. 
Switching off the running of the coupling, one 
cannot reproduce the behavior of high order cumulants. 
By relaxing the absolute normalization, thus building an effective model with 
one more parameter for each cumulant to reproduce the high energy region, 
 one  can  describe the experimental data only in a small  energy
region\cite{mont}. 
These results show the sensitivity of this analysis 
to the running of the coupling. In this respect, the moment analysis 
 of spectra in high-$p_T$ jets 
in $p\bar p$ collisions, where larger values of jet energy can be 
reached\cite{korytov}, is eagerly awaited.

\section{The invariant density in the soft limit} 

\subsection{A new scaling law in experimental data}

Let us consider the behaviour  of 
the charged particle  invariant density, $E dn/d^3p$,  
at small particle energy $E$.  
Figure~(\ref{charged}a)  shows data  at different $cms$ energies 
ranging from 3 GeV up to LEP-1.5\cite{klonew}.
The value of 270 MeV has been used for the effective mass  $Q_0$ 
which enters in the kinematical relation, $E^2 = p^2 + Q_0^2$.  
It is remarkable that 
all data scale  with $cms$ energies within 20\%  
at particle energy of the order of few hundreds MeV; 
at larger particle energies, a violation of the scaling-law is visible. 
LEP data seem to tend to a larger limiting value; it is not yet clear 
whether this is a signal of a different physics at LEP energies, like for
instance an enhanced contribution of weak decays to particle production in 
the soft region,  or a systematic effect in the overall normalization 
 of the different experiments.

\begin{figure}
\vfill \begin{minipage}{.48\linewidth}
          \begin{center}
   \mbox{
\mbox{\epsfig{file=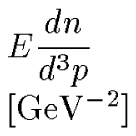,bbllx=5.2cm,bblly=17.5cm,bburx=5.4cm,bbury=25.cm}}
\mbox{\epsfig{file=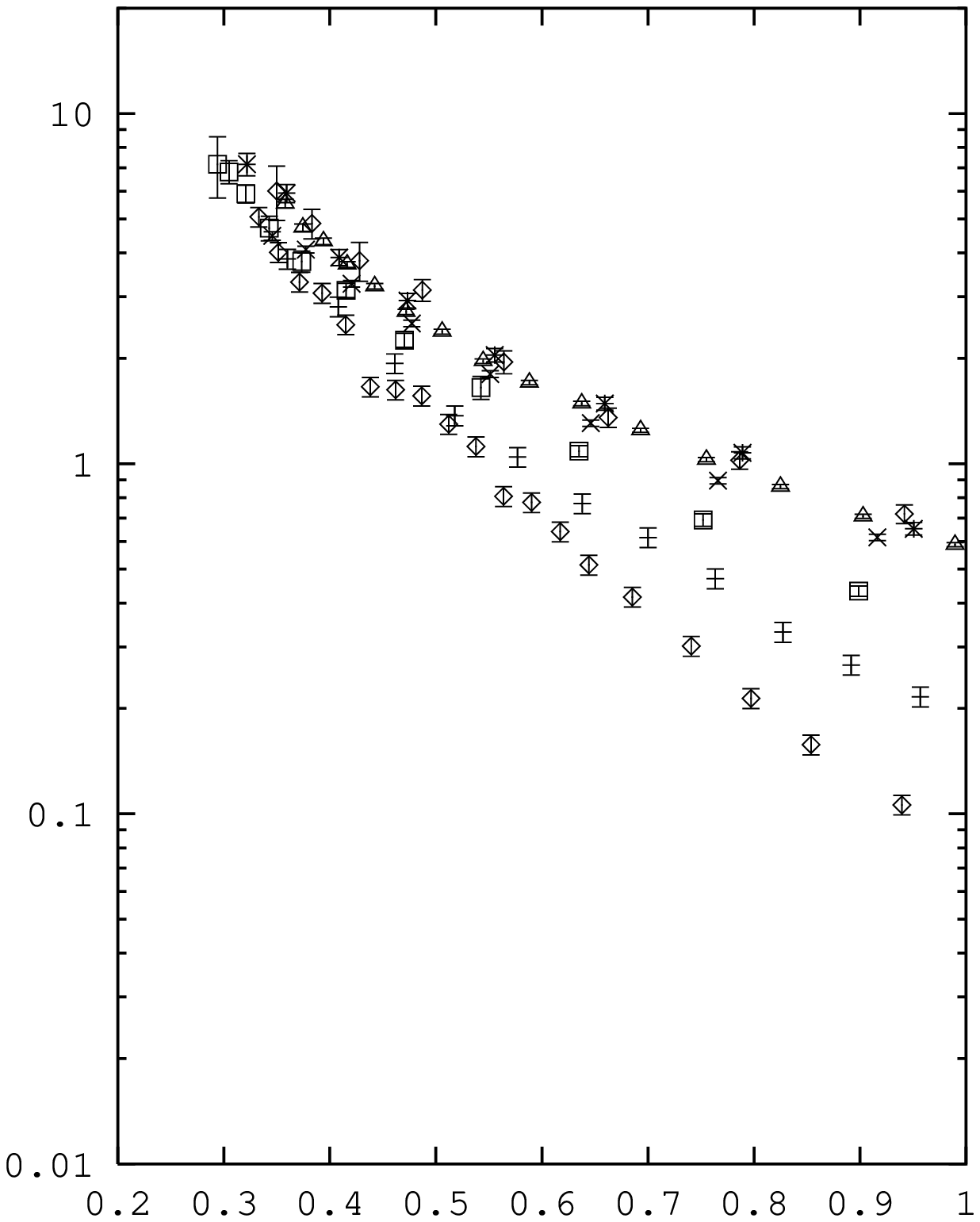,width=.72\linewidth,bbllx=5.5cm,bblly=2.5cm,bburx=15.cm,bbury=18.cm}}
  }        \end{center}
\vspace{-0.2cm}
\centerline{$\quad \qquad E$ [GeV]}
      \end{minipage}\hfill
      \begin{minipage}{.48\linewidth}
          \begin{center}
   \mbox{
\mbox{\epsfig{file=scritte.ps,bbllx=5.2cm,bblly=17.5cm,bburx=5.4cm,bbury=25.cm}}
\mbox{\epsfig{file=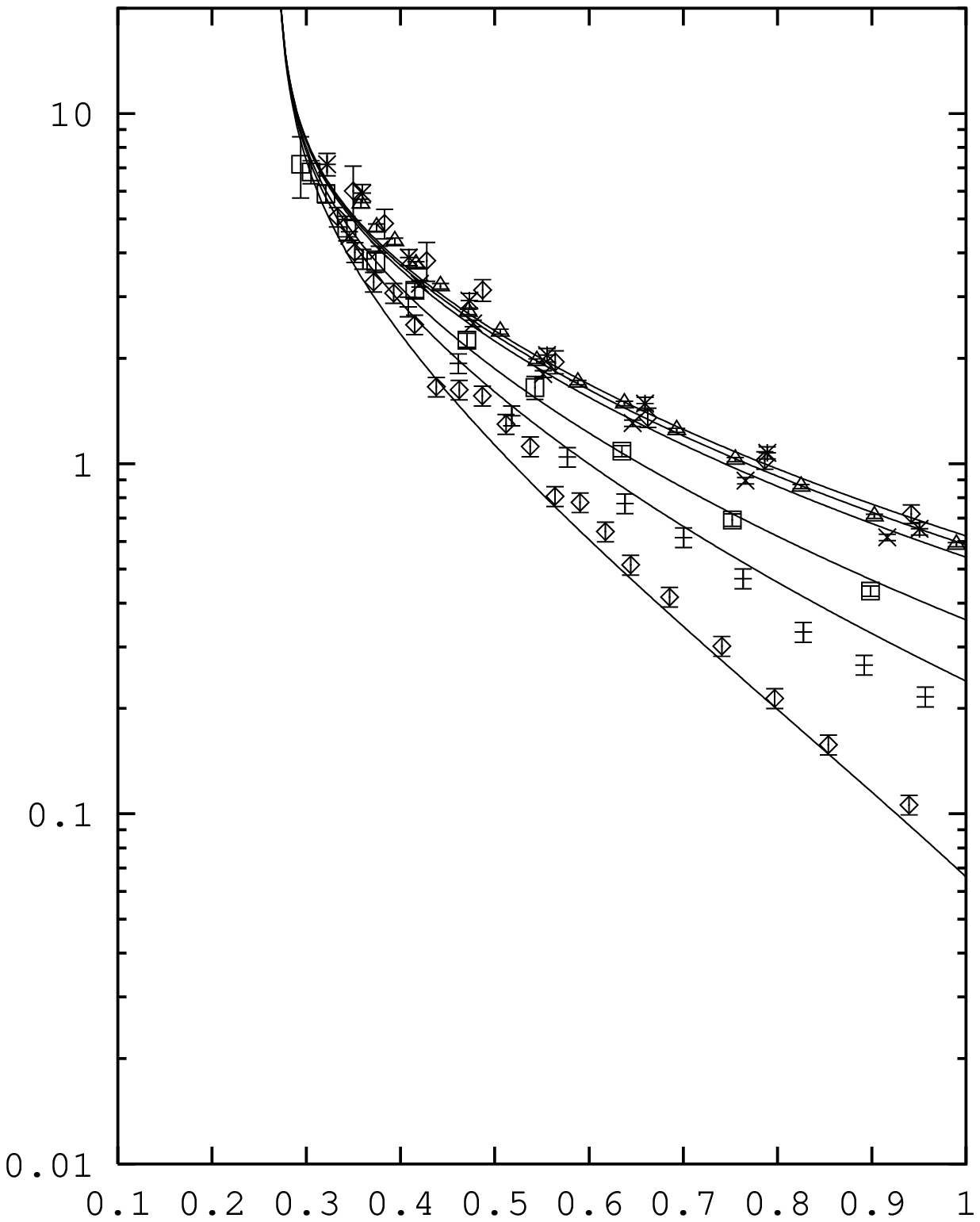,width=.72\linewidth,bbllx=5.5cm,bblly=2.5cm,bburx=15.cm,bbury=18.cm}}
  }        \end{center}
\vspace{-0.2cm}
\centerline{$\quad \qquad E$ [GeV]}
      \end{minipage}
\caption{{\bf a)}: Invariant density $E dn/d^3p$ for charged particle 
as a function of the particle energy $E$ with $Q_0$ = 270 MeV. 
{\bf b)}: same as in {\bf a)}, with theoretical predictions from MLLA with
the same $Q_0$ and $\Lambda$ = 257 MeV ($K_h$ = 0.45).} 
\label{charged}
\end{figure}

\subsection{Predictions of QCD coherence}

QCD colour coherence forbids branchings of very soft gluons\cite{DKMTbook}; 
therefore, the emission of soft partons should be proportional to the colour
charge of the initial system only and  the invariant
density $E dn/d^3p$ should  be independent of $cms$ energy 
at low particle energy\cite{vakcar}. 
Recently, predictions for the invariant density have been analytically 
computed\cite{lo,klonew} by solving explicitly in the limit of small particle energy
the evolution equation for 
the energy spectrum in a jet of type $A$, $D_A(\xi,Y,\lambda)$.  
The relation 
$E_h dn/d^3p_h \equiv K_h D_A(\xi_E,Y,\lambda)/[4\pi (E_h^2-Q^2_0)]$,  
with $E_h=\sqrt{p_h^2+Q_0^2}=E_p$, $\xi_E \equiv \log 1/x_E = \log \sqrt{s}/(2
E)$  and $K_h$ the overall normalization parameter, has been then 
used to obtain the invariant density $E dn/d^3p$. 

Within DLA, one gets an iterative solution for a gluon-jet\cite{lo}:  
\be
D_g(\xi,Y,\lambda) =  \delta(\xi) + 
\frac{4 N_C}{b} \log \left( 1 + \frac{Y-\xi}{\lambda}  \right) 
\left[ 1 + f(\xi,Y,\lambda) \right] + \dots
\ee
where $Y = \log \sqrt{s}/(2 Q_0)$, $\lambda = \log Q_0/\Lambda$, $b = 
\frac{11}{3}N_C-\frac{2n_f}{3}$   and 
$f(\xi,Y,\lambda)$ is a known function.  
The leading term at small momenta of order $\beta^2$
 corresponding to the emission of a single 
gluon  is  indeed proportional 
to the colour charge factor of the primary parton and  
does not depend on the $cms$ energy. The DLA prediction for the invariant 
density exhibits therefore an approximate scaling law and tends at 
small particle energy to a finite soft limit. 

The MLLA solution differs from the DLA limit by a simple 
exponential damping factor\cite{klonew}: 
\be
D(\xi,Y,\lambda)|_{MLLA} = D(\xi,Y,\lambda)|_{DLA} \exp \biggl[-a\int^Y_\xi
\frac{\alpha_s(y)}{2 \pi} dy \biggr] 
\ee
with $a=11/(3 N_C)+ 2n_f/(3N_C^2)$. 
This  solution satisfies the  MLLA evolution equation, 
except for a small term 
proportional to $a[\alpha_s(\xi)- \alpha_s(Y)]/(2 \pi)$, 
which then vanishes in the soft limit, where $\xi \to Y$. 
The MLLA still satisfies an approximate scaling law and tends to a finite 
soft limit, but the additional damping factor modifies the dependence on 
the $cms$ energy which appears at moderate particle energy. 
Figure~(\ref{charged}b) compares the data shown in 
Figure~(\ref{charged}a) with the theoretical predictions from MLLA  with 
$Q_0$ = 270 MeV and $\Lambda$ = 257 MeV (and $K_h$ = 0.45 fixed a posteriori); 
a rather good description of data is obtained within this approach.

\section{Conclusions}

The invariant density $E dn/d^3p$ has been shown to satisfy an approximate
scaling law at low particle energy of few hundreds MeV. 
In the framework of the perturbative QCD approach, 
which has been shown to successfully describe the invariant energy spectra 
in the whole available  $cms$ energy range, 
predictions for the invariant density $E dn/d^3p$ have
been explicitly computed. The MLLA has been found to describe well 
the data down to very low particle energies, 
thus suggesting  the validity of this picture 
in the very soft region and a possible link of the observed scaling law with
QCD coherence. 

Further tests of the universality of soft particle production in different
reactions and  alternative methods to point out 
the sensitivity of soft particle production to QCD coherence are presented 
in~\cite{klonew}.

 
\medskip
\noindent { \bf Acknowledgements} 
\medskip

I thank Valery A. Khoze and Wolfgang Ochs for discussions  
and collaboration on the subjects of this talk. 
I thank Jorge Dias de Deus 
for the nice atmosphere created at the Conference. 

\medskip
\noindent { \bf References}

\end{document}